\documentclass[sigconf]{acmart} %

\renewcommand\footnotetextcopyrightpermission[1]{}

\setcopyright{none}
\copyrightyear{none}
\acmYear{none}
\acmDOI{}

\acmConference[]{}{}{}

\usepackage{tikz}
\usepackage{amsmath}
\usepackage{listings}
\usepackage{multicol}
\usepackage{caption}
\usepackage{subcaption}
\usepackage{files/xurl}
\usepackage{dblfloatfix}
\usepackage{enumitem}  %

\definecolor{ckeyword}{HTML}{7F0055}
\definecolor{ccomment}{HTML}{3F7F5F}
\definecolor{cstring}{HTML}{2A0099}

\definecolor{aliceblue}{rgb}{0.86, 0.90, 0.95}

\lstdefinelanguage{Scala}%
{morekeywords={abstract,%
  case,catch,char,class,%
  def,else,extends,final,finally,for,%
  if,import,implicit,%
  match,module,lazy,%
  new,null,undefined,%
  override,%
  package,private,protected,public,%
  for,public,return,super,%
  this,throw,trait,try,type,%
  val,var,%
  with,while,%
  let,skip,assert,then,fst,snd,root,idx,sum,prod,exists,forall,%
  yield%
  },%
  sensitive,%
  escapechar=^,
  morecomment=[l]//,%
  morecomment=[s]{/*}{*/},%
  morestring=[b]",%
  showstringspaces=false%
}[keywords,comments,strings]%

\definecolor{listingbg}{RGB}{240, 240, 240}

\newcommand{\commentstyle}[1]{\color{ccomment}\itshape{#1}}
\newcommand{\keywordstyle}[1]{\color{ckeyword}\bfseries{#1}}
\newcommand{\stringstyle}[1]{\color{cstring}\bfseries{#1}}

\lstnewenvironment{listing}{\lstset{language=Scala}}{}
\lstnewenvironment{listingtiny}{\lstset{language=Scala,basicstyle=\scriptsize\ttfamily}}{}

\renewcommand{\d}[1]{\sfrac{d\!}{d #1}\ }

\newcommand{\silent}[1]{}

\begin{document}

\date{}

\title{Architecting Intermediate Layers for Efficient Composition of Data Management and Machine Learning Systems}

\author{Supun Abeysinghe, Fei Wang, Gregory Essertel, Tiark Rompf}
\affiliation{
  \institution{Purdue University}
  \city{West Lafayette}
  \state{Indiana}
  \country{USA}
}
\email{{tabeysin, wang603, gesserte, tiark}@purdue.edu}

\begin{abstract}

Modern data analytics workloads combine relational data processing with machine learning (ML). 
Most DBMS handle these workloads by offloading these ML operations to external specialized ML systems. 
While both DBMS and ML systems go to great lengths to optimize performance for their specific workloads,
significant performance is lost when used in combination, due to data movement across system boundaries,
conversions between incompatible internal data formats, and the lack of cross system optimizations. 

A key idea to remove these bottlenecks is to integrate existing data manipulation systems with ML 
systems by building a common intermediate layer (IR). Although this idea has been explored before 
(Weld, Delite), previous such attempts require significant re-engineering of prior systems and still
fall short in achieving best-of-breed performance for individual tasks (e.g., SQL, Deep Learning). 
Specifically, they rely on re-implementing existing systems using a generic set of operators and 
fail to match best-of-breed individual performance due to the inability to recover high-level 
optimizations from this generic IR through compiler analysis.

We present Flern, the first intermediate-layer integration between DB and ML systems that are
best-of-breed individually, competitive with the best compiled query engines such as HyPer on
comprehensive relational benchmarks (TPC-H) and competitive with TensorFlow and PyTorch in state-of-the-art
ML models (e.g., DeepSpeech, SqueezeNet, Transformers) and also represents a new state-of-the-art for
integration. A key realization is to architect intermediate layers based on generative programming
capabilities, which preserves high-level contextual information for cross optimizations and enables
the construction of a variety of complex structures and cross system optimizations with minimal effort.
\end{abstract}

\settopmatter{printfolios=true}
\maketitle %
\pagestyle{plain}

\section{Introduction}
\label{section:intro}
Today's data analytics workloads often combine traditional relational processing with modern machine learning
(ML) operations. These two workload types are significantly different from each other and have different
domain-specific properties that lead to the construction of independent specialized systems (i.e., DBMS
for relational processing and ML frameworks for ML operations)~\cite{dbml_challenges_18}. These systems go to great lengths to
optimize their workloads by taking domain-specific properties into account (e.g., query compilation,
binding to low-level specialized kernels).

However, when complex data analytics pipelines are built by combining such systems, their end-to-end
performance becomes suboptimal. This can be due to multiple reasons, including the inability to perform
global optimizations across system boundaries and the overhead of data movement and conversion at these
boundaries, often caused by incompatible data formats. For instance, a simple relational query that uses
an ML model inside Postgres (using PyTorch for ML) can be more than 50x slower than a manual
hand-optimized program written for the same task.

Our work aims to address this key challenge: 
how can we efficiently manage workloads that combine relational processing with ML? 
Our objectives are threefold: 1) to eradicate costly overheads at system boundaries,
2) to perform global optimizations on the entire computation, rather than treating ML
computations as black boxes,
and 3) to achieve these goals without the need to re-engineer systems from the ground up.

One approach is to add support for ML computations inside DBMS by extending SQL engines with new
operators (e.g., user-defined functions, 
iterative computations, etc.)~\cite{dbml_db4ml, dbml_gpu, bismarck, factml_morpheus_1, factml_lmfao, madlib}.
However, these approaches
have limited support for the types of models they can run and do not support modern ML models 
(e.g., Transformers~\cite{DBLP:conf/nips/VaswaniSPUJGKP17}) due to lack of expressiveness.
Moreover, the performance is not on par with
specialized ML systems (e.g., PyTorch, TensorFlow, etc.) mainly due to the inability to incorporate
domain-specific optimizations.

Another, more intrusive but also more powerful, approach is to build a common intermediate layer across
different systems. The intermediate layer must be sufficiently generic to support all operations of disjoint
systems. That is, it should support all SQL/DataFrame operations (e.g., different types of joins,
aggregation queries, etc.)
and all deep learning operators (e.g., automatic differentiation (AD), complex ML models, etc.).
A line of work that follows this idea of building common intermediate
layers~\cite{DBLP:conf/cidr/PalkarTSSAZ17, DBLP:journals/pvldb/PirkMZM16, DBLP:journals/tecs/SujeethBLRCOO14}
achieves this by imposing a single, fixed intermediate layer as a \emph{one-size fits all}
consisting of a handful of generic operators (e.g., \emph{map}, \emph{reduce}, etc.).
This eliminates overheads at system boundaries by generating a unified executable.
While there is considerable flexibility for optimization, this generic IR lacks vital contextual
information which makes it difficult to recover low-level performance through compiler analysis.
Moreover, all the high-level operators
(e.g., Joins, Tensor operations, etc.) of existing systems (or their front-ends) should be re-implemented using
these generic IR constructs which require a significant engineering effort and can be challenging for complex
operators (e.g., in Section~\ref{section:motivation}, we investigate a Hash Join implementation of such a system
and show how it differs from a classic textbook implementation).

Another approach is constructing high-level IRs comprising high-level DB and ML ops.
For instance, Raven~\cite{raven_sigmod} extends ONNX~\cite{onnx}, an IR for ML, with
relational algebra operations to construct an IR tailored for combined workloads.
This strategy facilitates cross-workload optimization, as it provides access to
high-level abstractions and rich contextual information, as opposed to using a generic,
fixed IR.
While this approach can improve end-to-end performance by enabling cross-workload
optimizations, it still incurs significant overheads at system boundaries due to data copying
and format conversion.
This is because, despite creating a unified IR, separate runtime systems remain in use for DB
(such as Spark SQL~\cite{DBLP:conf/sigmod/ArmbrustXLHLBMK15} or SQL Server) and
ML (like ONNXRuntime~\cite{onnxruntime}) execution.

In this work, we investigate the potential for achieving the best-of-both worlds. That is, having the capability 
to perform general, reusable compiler optimizations using a low-level common IR and generating a unified executable 
that eliminates overheads at system boundaries and also, having access to high-level abstractions and contextual 
information to add cross-system optimizations. To achieve this goal, we leverage an existing relational query 
compiler, Flare~\cite{DBLP:conf/sigmod/TahboubER18, DBLP:conf/osdi/EssertelTDBOR18}, and an ML compiler, 
Lantern~\cite{DBLP:journals/pacmpl/WangZDWER19, DBLP:conf/nips/WangDWER18}, both of which are built using 
generative programming (introduced in Section~\ref{subsec:why_gp})~\cite{DBLP:conf/gpce/RompfO10}.
Specifically, we integrate them at a higher-level to handle combined relational and ML workloads.

Generative programming has demonstrated great effectiveness in architecting large-scale systems for 
individual workloads with relatively smaller engineering effort compared to highly engineered counterparts, 
while achieving competitive
performance~\cite{DBLP:conf/sigmod/TahboubER18, DBLP:conf/osdi/EssertelTDBOR18,gensym_oopsla20,sai_oopsla19}. 
This generative approach allows programmers to develop their systems using a high-level language with 
user-friendly features (e.g., high-level type system, high-level data structures, abstractions such as 
classes, interfaces, generics, etc.) while achieving native performance by translating this high-level 
code to low-level native code (thus, \emph{generative}) which does not contain any of the high-level 
abstractions, hence, \emph{abstraction without regret}~\cite{tr_thesis}.

In this work, we investigate the ways to reconfigure such
systems designed around the idea of generative programming
to efficiently compose multiple such systems to handle complex
workloads.
We demonstrate this generative programming approach enables
efficient post-hoc integration of such specialized systems
with relatively smaller engineering effort.
Specifically, we build Flern, a query
compiler that supports full ML capabilities, by integrating Flare~\cite{DBLP:conf/sigmod/TahboubER18, DBLP:conf/osdi/EssertelTDBOR18},
a state-of-the-art query compiler
and Lantern~\cite{DBLP:journals/pacmpl/WangZDWER19, DBLP:conf/nips/WangDWER18} an ML framework,
that are developed
based on the same generative programming technique called Lightweight Modular Staging (LMS)
\cite{DBLP:conf/gpce/RompfO10}. Flern is the first system that is best-of-breed individually
(competitive with state-of-the-art in-memory SQL engines like HyPer \cite{DBLP:journals/pvldb/Neumann11}
on full relational benchmarks \cite{DBLP:conf/osdi/EssertelTDBOR18} e.g., TPC-H  and competitive with
TensorFlow and PyTorch in sophisticated deep learning models
\cite{DBLP:journals/pacmpl/WangZDWER19} \cite{DBLP:conf/nips/WangDWER18} e.g., DeepSpeech
\cite{DBLP:conf/icml/AmodeiABCCCCCCD16}, SqueezeNet \cite{DBLP:journals/corr/IandolaMAHDK16},
Transformer \cite{DBLP:conf/nips/VaswaniSPUJGKP17}) and also the new state-of-the-art for integration.

\begin{figure}[t]
	\centering
	\includegraphics[width=60mm]{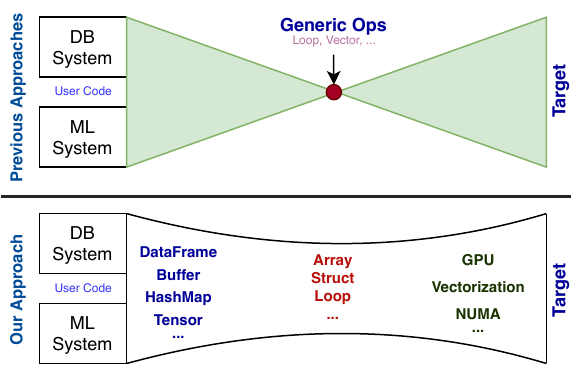}
	\caption{
	Previous approaches directly map the user program to a generic IR, lacking necessary contextual
	information for high-level global optimizations.
	In contrast, Flern preserves high-level abstractions and vital contextual information,
	enabling seamless introduction of cross-optimizations.
	}

	\vspace{-4mm}
	\label{figure:intro}
	
\end{figure}

\textbf{Contributions}
This paper's primary intellectual contribution lies in its examination of the constraints inherent
to existing specialized code generation methods when dealing with multi-paradigm workloads,
specifically DB and ML.
Then, we present an approach for tackling this challenge of system integration in general:
architect systems based on generative programming, so that they can be adapted more
effectively and at a lower engineering cost, 
and provide evidence of the approach's effectiveness by building Flern.
Our specific contributions are summarized as follows:
\vspace{-4mm}
\begin{itemize}[leftmargin=1em]
	\item We analyze the limitations of existing approaches to build common intermediate layers
	and discuss why generative programming emerges as a superior choice for integrating systems efficiently
	with minimal re-engineering cost (Section \ref{section:motivation})
	
	\item We demonstrate how generative programming enables efficient post-hoc integration of prior
	systems by combining two state-of-the-art systems in DB and ML domains. 
	Specifically, we illustrate how our approach can be utilized to eliminate overheads at
	system boundaries and implement cross-optimizations, all while maintaining the best-of-breed
	performance of individual systems.
	(Section \ref{section:overview})
	
	\item We present two sets of benchmarks where we first evaluate the performance impact of
	each introduced cross-optimization, followed by a comparison of the performance against
	state-of-the-art baselines. This comparison demonstrates that our approach 
	(1) either outperforms or achieves competitive results in isolated tasks
	(i.e., data manipulation and ML independently), and 
	(2) delivers state-of-the-art performance—demonstrating up to an
	order of magnitude speedups when applied to combined tasks (Section \ref{section:experiments}).

\end{itemize}

In Section~\ref{section:related}, we present an analysis of related work in this domain.
Finally, in Section \ref{section:conclusions}, we draw conclusions and discuss potential
future research directions.
A preliminary demonstration paper outlining the initial conception and operation
of our proposed system was previously published~\cite{flern_vldb_demo}.
In this current work, we add further optimizations on top of the initial integration,
and conduct a comprehensive experimental evaluation, comparing our system against
well-established baselines.

\section{Motivation}\label{section:motivation}

\subsection{Why Intermediate Layer Integration?}

Specialized ML systems commonly employ various techniques to achieve optimal performance,
such as the usage of high-level abstractions (e.g., Tensor) that make it relatively
easy to add domain-specific optimizations (e.g., Tensor transformations). 
To achieve performance parity with these specialized ML systems during mixed
DB and ML workloads,
the capability to incorporate these optimizations becomes vital.
However, this task can prove challenging when ML operations are incorporated into DBMS
using divergent abstractions and methodologies (e.g., extended SQL execution engines, 
user-defined functions, etc.).
The key hurdle is the intricacy involved in adapting the existing optimization strategies
to these different settings.

\begin{figure}[t!]
\begin{lstlisting}[language=Python,basicstyle=\fontsize{6.5}{1}\selectfont\ttfamily]
def join(expr1, expr2, d1_keys, d2_keys, keys_type,  ..):
	""" Computes a join on two tables """
	weld_obj = WeldObject(encoder_, decoder_)
	
	df1_var = weld_obj.update(expr1)
	if isinstance(expr1, WeldObject):
		df1_var = expr1.obj_id
		weld_obj.dependencies[df1_var] = expr1
	df2_var = weld_obj.update(expr2)
	if isinstance(expr2, WeldObject):
		df2_var = expr2.obj_id
		weld_obj.dependencies[df2_var] = expr2
	#Some String manipulations constructing holes of the template elided
	weld_template = """
	let df2_join_table = result(
		for(
			groupmerger[%
			|b, i, e| merge(b, {%
		)
	);
	result(for(
		appender,
		|b, i, e|
			for(
				lookup(df2_join_table, %
				b,
				|b2, i2, e2| merge(b, {%
			)
	))"""
			
	weld_obj.weld_code = weld_template %
		
	return weld_obj
\end{lstlisting}
\caption{Join Implementation of Weld (for accelerating Pandas). The operators emit
code for IR construction as blocks of strings. This kind of manipulation of code in
stringified form is generally error-prone, relatively harder to maintain and implement.}
\label{code:weld_join}
\vspace{-4mm}
\end{figure}

An alternative strategy is to integrate a DBMS with an existing ML system already furnished
with essential domain-specific optimizations.
Such integration potentially curtails the need for substantial re-engineering efforts.
A case in point is the PL/Python plugin~\cite{plpython}, which facilitates the execution of
Python-based arbitrary procedures within PostgreSQL.
As a result, any ML framework featuring Python interfaces, such as PyTorch~\cite{DBLP:conf/nips/PaszkeGMLBCKLGA19}
or TensorFlow~\cite{DBLP:conf/osdi/AbadiBCCDDDGIIK16}, can be employed within the database for
ML operation implementation.
In this way, the DBMS essentially gains full access to a fully-fledged ML system's capabilities for 
effectively managing combined relational and ML workloads.

This approach, however, encounters a significant drawback.
Both systems tend to operate as separate entities, which leads to data copying, format
conversions at system boundaries, and hinders cross-system optimizations.
For instance, in the Postgres scenario, data must be transferred from the database to
the Python environment, which then needs to be converted into PyTorch's tensor data format.
Once the ML computation concludes, these tensors have to be transformed back into PL/Py objects,
copied back to Postgres, and finally converted into the original record format.
For example, when executing a three-layer neural network model on the NYC-Taxi dataset~\cite{nyc_dataset}
using this approach, the total execution time comes to 638 seconds whereas time spent on doing
actual ML computations comes to around 6 seconds, underlining the substantial overheads
involved (examined in detail in Section~\ref{section:eval_ml_udf}).

To mitigate or even eradicate these overheads, we turn our attention to query
compilers~\cite{DBLP:conf/osdi/EssertelTDBOR18, DBLP:conf/sigmod/TahboubER18, DBLP:journals/pvldb/Neumann11}.
These compilers accelerate query execution speed by generating specialized, low-level code
tailored to a particular query.
Generally, these systems build an IR for the incoming query, conduct a series of optimizations
on the IR, and then generate code from the optimized IR.
This generated code is compiled by a generic compiler and executed to produce
the final result.
One strategy to eliminate the data movement overheads across systems in the combined 
workload setting is to combine the systems at this IR level.
This entails lowering both the query compiler and the ML computations to a single common IR,
enabling global optimizations and minimizing data movement overheads by 
carefully analyzing the IR and generating a unified executable.

\vspace{-2mm}
\subsection{Why Generative Programming?}\label{subsec:why_gp}

\begin{figure}[t!]
	\begin{lstlisting}[language=Scala,basicstyle=\fontsize{6.5}{5}\selectfont\ttfamily]
		case class HashJoinOp(left: Op, right:Op)(/* elided */) extends Op {
			// HashMap specialized for corresponding Schemas
			val map = LinkedHashMap(keySchema, left.schema)
			def exec(callback: Record => None) = {
				// Store records from left operator tree in the HashMap
				left.exec { tuple =>
					map.update(leftHash(tuple), tuple)
				}
				
				// Retrive records from right subtree and join
				right.exec { tuple =>
					for (lTuple <- map(rightHash(rTuple)) if joinCond(lTuple, rTuple))
					callback(lTuple ++ rTuple)
				}
			}
		}
	\end{lstlisting}
	\caption{Hash-Join Implementation of Flare which is a query compiler based on generative programming.
	The implementation contains the operator logic implemented using a high-level programming language (Scala)
	as opposed to emitting blocks of stringified code and has access to all the features of the host language
	(e.g., type system, abstractions, etc.).}
	\vspace{-5mm}
	\label{code:lb2_join}
	
\end{figure}

Delite~\cite{DBLP:journals/tecs/SujeethBLRCOO14} and Weld \cite{DBLP:conf/cidr/PalkarTSSAZ17} are 
prior systems that follow the strategy of developing common IRs across multiple systems,
which allows for the generation of unified executables.
Both of these methods share a similar characteristic in utilizing a fixed, generic intermediate layer
composed of a limited set of generic operators that front-ends use to implement their functionalities.
This offers substantial flexibility as any optimizations added to this generic IR layer can be
reused by any front-end.
However, this flexibility tends to come with notable trade-offs.
Firstly, the generic nature of the IR means that most of the contextual information originating from
high-level abstractions such as \texttt{tensors} or \texttt{DataFrame} will be lost at the IR level.
For example, if a 2D tensor is transposed twice consecutively, it is relatively easy to determine that this
results in a no-op if we have access to contextual information about the data (i.e., that it is a 2D tensor),
but this would require compiler analysis of the looping structures in the generic IR.
Secondly, implementing complex operators using these minimal IR constructs can pose a challenge.
This is because their actual implementations may significantly deviate from their textbook versions.

For instance, Weld defines a minimal IR that encapsulates the structure of prevalent data-parallel
algorithms, complemented by a runtime API that allows disjoint libraries to construct Weld IR fragments.
These host libraries need to fully re-implement their operators to emit
the necessary IR fragments that reflect the desired operator logic.
The backend then accumulates these emitted IR fragments into a unified global IR.
This combined IR undergoes several stages of optimization and drives the process of
generating optimized code via LLVM~\cite{DBLP:conf/cgo/LattnerA04}

Figure~\ref{code:weld_join} shows the Weld implementation of the \texttt{Join} 
operator~\cite{DBLP:conf/cidr/PalkarTSSAZ17}.
Notably, the implementation deviates substantially from a conventional textbook implementation
and consists of a stringified code template for the \texttt{Join} operator containing 
placeholders that are filled based on function parameters.
However, since this kind of template expansion is
limited by the fact that the code is manipulated as strings, it is either impossible or challenging
to use features of high-level programming languages, such as type-checking, on these code templates.
As a result, these code templates often pose challenges in implementation and maintenance.
Consequently, systems adopting this method usually support a restricted set of cases
(e.g., a single type of join, or a limited set of ML operations) due to the engineering effort
required and the complexity involved.
Therefore, while there's potential for improvement in end-to-end performance, these approaches
often either lack support or fail to reach best-of-breed performance for specific workloads
in comprehensive benchmarks (such as the full TPC-H benchmark).

In contrast, generative programming is fine-grained, supports lots of programmatic control,
and leverages rich contextual information.
Take Lightweight Modular Staging (LMS)~\cite{DBLP:conf/gpce/RompfO10}, for example. 
It's a generative programming framework in Scala that is based on the idea of 
multi-stage programming~\cite{DBLP:journals/tcs/TahaS00} (commonly known as staging).
Multi-stage programming is a paradigm that equips developers with the ability to write
generic programs using higher degrees of abstractions without compromising on 
runtime efficiency~\cite{DBLP:conf/gttse/Taha07}. 
The high-level idea behind staging is to delay
computations of certain operations to a later stage, thereby generating code for such operations
with the information known in the current stage.
An illustrative example of LMS in action is depicted in the following code snippet.
Note that the implementation closely mirrors programming in a conventional language,
compared to resorting to manipulating large blocks of stringified code templates.

\begin{lstlisting}[language=Scala,basicstyle=\fontsize{8.5}{5}\selectfont\ttfamily\footnotesize]
def power(b: Rep[Int], n: Int):Rep[Int] =
	if (n == 0) 1
	else if (n %
		power(b, n/2) * power(b, n/2)
	else b * power(b, n-1)
def main(args: Rep[Array[String]]) =
	println(power(args(0).toInt, 7))
\end{lstlisting}

LMS is type-driven, and utilizes \texttt{Rep} types to denote values that will be
computed in the subsequent stage, and therefore, should be computed by the generated code
(\texttt{Rep} \textbf{rep}resents next-stage expressions).
All operations, including language constructs such as control flow on conventional 
non-\texttt{Rep} expressions (for instance, \texttt{Int}), are evaluated at the present stage.
In contrast, all \texttt{Rep}-typed expressions contribute to code generation.
In the given code snippet, \texttt{b} is of a \texttt{Rep}-type,
denoting it is not known in the current stage, while \texttt{n} is of a
standard type, thus, its value is known.
This allows for partial evaluation of the program based on
\texttt{n}, generating code for later stage execution.
Moreover, LMS is capable of directly generating equivalent
low-level C code for the partially evaluated Scala code (for performance reasons).
Below is the resultant C code.

\begin{lstlisting}[language=C,basicstyle=\fontsize{8.5}{1}\selectfont\ttfamily\footnotesize]
int main(int argc, char** argv) = {
	int x0 = atoi(argv[0]); // args(0).toInt
	int x1 = x0 * (x0 * x0); // x1 = x0 ** 3
	printf("%
	// x0 * ((x0 ** 3)* (x0 ** 3)) = x0 ** 7
}
\end{lstlisting}

A notable observation is that the original source code, written in Scala,
closely mirrors typical Scala code, except for the type annotation which
indicates the variables that needs to be staged.
This pattern generally holds true across most instances,
implying that a developer can transform a standard Scala program into a
staged program primarily by altering the types.

Under the hood, LMS maintains an extensible graph-like IR to capture the constructs
and operations of the staged program.
It features numerous IR-level optimizations (including loop fusion, common 
sub-expression elimination, loop unrolling, and function inlining), 
and the IR's extensible nature allows library developers to conveniently add their
domain-specific optimizations~\cite{lmsopt}.
In this context, we can perceive LMS as enabling us to utilize the entirety of
Scala as a macro language in the construction of the intermediate layer.
This consequently simplifies the implementation of intricate data structures and
algorithms required for sophisticated data manipulations and comprehensive
deep learning functionalities.
Furthermore, LMS has the ability to generate low-level code that does not contain
any high-level abstractions utilized in the original 
Scala program (e.g., \texttt{Tensor}, \texttt{DataFrame}), 
thereby eliminating potential runtime penalties tied to these abstractions.

Prior work has demonstrated the effectiveness of generative programming as a
systems-building technique that dramatically simplifies the construction of
high-performance systems that have a high degree of internal variability and need
specialization to achieve performance (e.g., Flare~\cite{DBLP:conf/osdi/EssertelTDBOR18},
LB2~\cite{DBLP:conf/sigmod/TahboubER18}). For example, Figure~\ref{code:lb2_join} shows
the implementation of \texttt{HashJoin} operator in LB2. 
This implementation equates to writing the hash join algorithm for a straightforward query
interpreter, essentially following textbook pseudo-code, using a high-level programming language,
in this case, Scala. 
The developer has full access to high-level data structures (e.g., \texttt{LinkedHashMap})
specialized for the given schemas, abstractions (e.g., \texttt{Record}), and the Scala type system
as opposed to writing \textit{brittle} stringified code templates.

In this work, we demonstrate that 
in the case of composing such systems by constructing common IRs, these properties of
generative programming are useful for adapting system boundaries with minimal overhead
and implementing key global optimizations that span across system boundaries.
It permits the implementation of such optimizations using varied abstraction
levels (see Figure~\ref{figure:intro}) while utilizing a high-level programming language,
with complete access to contextual information associated with high-level operators.
Essentially, this implies that cross-optimizations can be implemented using
high-level abstractions like \texttt{Tensor} and \texttt{DataFrame},
as opposed to mandating all global optimizations to be carried out on a generic IR.
In Section~\ref{section:overview}, we illustrate how these generative programming
properties can be harnessed to seamlessly combine a generative programming-based
query compiler with a ML framework incurring 
minimal overheads at system boundaries and how we can
implement cross system operations and optimizations to achieve state-of-the-art
performance in combined DB and ML workloads.

\vspace{-2mm}

\section{Flern Overview}\label{section:overview}

\begin{figure}[t!]
	\centering
	\includegraphics[width=0.30\textwidth]{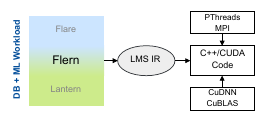}
	\caption{Flern's architecture: Seamlessly integrating Flare and Lantern at a common IR level and
	adding cross-system optimizations at the boundary.
	It generates a unified executable for combined DB and ML workloads.}
	\vspace{-2mm}
	\label{figure:overall}
	
\end{figure}

In this section, we illustrate how the benefits of generative programming can be
harnessed to seamlessly integrate two pre-established systems from the realms of DB (relational)
and ML (tensor) for the processing of combined workloads.
Specifically, we build Flern using Flare~\cite{DBLP:conf/osdi/EssertelTDBOR18}, 
an SQL query compiler, and Lantern~\cite{DBLP:journals/pacmpl/WangZDWER19, DBLP:conf/nips/WangDWER18, DBLP:conf/iclr/WangR18},
an ML framework (Figure~\ref{figure:overall}).
Both of these systems have been developed based on the same generative programming approach, 
Lightweight Modular Staging (LMS)~\cite{DBLP:conf/gpce/RompfO10}.
This demonstrates how Flern can benefit from the distinct strengths of these two systems,
which each offer full functionality in their respective domains, such as comprehensive SQL
support and the capability to handle modern deep learning models.
Moreover, both Flare and Lantern deliver performance that competes effectively with
the best-in-class systems in their respective fields.

\begin{figure}
	\lstset{basicstyle=\ttfamily\fontsize{7.2}{8.64}\selectfont, columns=fullflexible}
	\begin{lstlisting}[language=Python]
# Flern: ML in-charge: training models by querying DB for data
for epoch in range(num_epochs):
  for batch, target in sql("select ... from t1 join t2 ..."):
    model.train(batch, target)

# Flern: DB in-charge: using ML models in SQL queries
sql.register_udf("classifier", model)
sql("select p, classifier(xs) from r;")
	\end{lstlisting}
	\caption{Flern's main use cases: ML model training with 
	DB data retrieval using SQL queries, and DB query evaluation
	with trained ML models in relational queries.}
	\label{fig:flern_cases}
	\vspace{-3mm}
\end{figure}

\textbf{Flare:}
Flare is built on top of LB2 \cite{DBLP:conf/sigmod/TahboubER18}, a high-level query compiler developed
using generative programming techniques. LB2 implements relational operators in a way that is similar to
a simple query interpreter and converts it to a query compiler using Futamura projections and partial
evaluation \cite{futamura}. Flare operates on the optimized query plans generated by Spark's query
optimizer (Catalyst \cite{DBLP:conf/sigmod/ArmbrustXLHLBMK15}) derived from user queries and performs
query compilation and runtime native code generation. This code generation strategy of Flare achieves
orders of magnitude speedups over Spark and other RDBMS and achieves competitive performance with
state-of-the-art in-memory query compilers like Hyper \cite{DBLP:journals/pvldb/Neumann11} in full
relational benchmarks (e.g., TPC-H) \cite{DBLP:conf/osdi/EssertelTDBOR18}.

\textbf{Lantern:}
Lantern~\cite{DBLP:conf/nips/WangDWER18, DBLP:journals/pacmpl/WangZDWER19, DBLP:conf/iclr/WangR18}
is a differentiable programming framework that handles automatic differentiation via delimited
continuations~\cite{DBLP:journals/lisp/Fischer93}, and code generation via LMS \cite{DBLP:conf/gpce/RompfO10}.
Delimited continuations allow Lantern to support ML models with in-graph control-flows such as conditionals,
loops, and functions. LMS reifies the computation graph (after automatic differentiation) for code generation
that utilizes various BLAS and neural network kernel libraries for multiple hardware platforms
(such as CPU and GPU). Lantern performs competitively with existing deep learning frameworks
(e.g., PyTorch and TensorFlow) in modern deep learning models.

Figure~\ref{fig:flern_cases} illustrates the two main scenarios where
Flern can be utilized.
Generally, combined DB and ML workloads can be classified into two major
categories.
The first category encompasses cases where ML models are trained using
data obtained through DB queries.
This also includes end-to-end data science pipelines constructed using data
manipulation tools (e.g., Pandas~\cite{pandas}, Spark) and ML frameworks
(e.g., PyTorch, TensorFlow).
In these scenarios, the ML system repeatedly retrieves data in batches from the
data manipulation system for model training.
The second category involves instances where queries utilizing pre-trained ML models 
are executed on a DB. For instance, an SQL query may contain one or more pre-trained
ML classifiers as functions.

Recall that our primary design objective is to extend DBMS with ML capabilities with
minimal reconfigurations to existing systems, ensuring they maintain their individual
performance while improving their combined performance.
Flare and Lantern achieves competitive performance with state-of-the-art systems for their respective workloads.
Hence, one straightforward approach would be to operate these two systems independently
and transfer data between them using a commonly used data format, such as CSV.
This approach wouldn't require significant modifications to either system and
could be implemented by introducing a CSV exporter and importer at the respective
system boundaries.
While this approach would maintain the performance of individual systems,
it introduces a substantial overhead due to the necessity of exporting data
in a chosen format, loading this data, parsing the loaded data, etc.
(as evaluated in Section~\ref{section:individiual_optimizations}).

A modest improvement on this naive implementation could involve using a
binary format to decrease the overhead associated with data formatting and parsing,
and to facilitate communication over memory (e.g., using pipes or shared memory).
However, communication overheads persist due to the separation of runtime environments.
Furthermore, in scenarios where complex relational queries are mixed with tensor
computations (e.g., ML Classifiers as UDFs), this data movement may need to occur
multiple times in both directions, necessitating careful modifications to both systems
to handle such interactions with external systems.

\vspace{-2mm}
\subsection{Adapting the Boundaries of Flare and Lantern}\label{subsec:adapting_boundary}

A crucial characteristic of both Flare and Lantern is that they are built upon
the same generative programming paradigm - Lightweight Modular Staging (LMS).
As discussed in Section \ref{section:motivation}, a key benefit of this generative
strategy is that these high-performance systems are implemented in Scala, a
feature-rich high-level high-level programming language.
Consequently, we can combine them simply by importing them as Scala libraries,
granting us access to the functionalities of both systems.

However, instead of combining them using the conventional method of transferring data
across respective library functions, which incurs data movement and conversion overheads,
we propose bringing the two systems together at a common layer and constructing a single
IR graph at the LMS level.
More specifically, we devise a shared layer that supports the union of operators from
both systems while preserving all the transformations and optimizations inherent to
each individual workload.

The feasibility of this approach owes much to the modular architecture of both systems
(and systems architected in Scala in general).
Specifically, functionalities are enclosed in Scala traits as discrete modules,
and these more compact modules are combined using class compositions with 
mixins~\cite{mixin} to construct a larger system.
For instance, Flare defines traits such as \texttt{CompileProject},
\texttt{CompileJoin}, and \texttt{CompileAggregate} to manage the operator
logic for Project, Join, and Aggregate operations.
These smaller building blocks are then combined into \texttt{FlareOps}.
Lantern also employs a similar design strategy.
Thus, by leveraging this modular structure, we can integrate the functionalities
of both systems using a similar mechanism.
We then tailor the LMS backend so that it builds a single IR graph containing
operations from both workloads.

Despite their compatibility, the two systems operate using distinct high-level
abstractions, and they are not aware of each other's constructs (e.g., Flare does
not recognize the \texttt{Tensor}s in Lantern, and vice versa with \texttt{Record},
\texttt{Buffer}, etc., in Flare).
As a result, we must facilitate conversions between these high-level abstractions at
the boundaries of each system.
However, a distinctive trait of generative programming-based code generation is that
these high-level abstractions dissolve into native data structures in the generated
code (for instance, \texttt{Tensor} and \texttt{Buffer} transform into
native C arrays).
Consequently, even though we carry out conversions between these abstractions
in the high-level code, these conversions effectively become simple variable
assignments in the generated code.
This characteristic ensures that there is no adverse impact on performance
during runtime.

Thanks to the inherent extensibility of
LMS~\cite{lms_reflections, lmsopt, DBLP:conf/gpce/RompfO10}, implementing
these conversions is a relatively straightforward task.
For instance, to transform a \texttt{Buffer} into a \texttt{Tensor}, we can define
a \texttt{toTensor} method that establishes a new node type
(e.g., \texttt{to-tensor}) in the IR.
Subsequently, we can implement the code generation logic for this newly established
node, which translates to a trivial variable assignment in the
generated code.
Alternatively, considering that both \texttt{Buffer} and \texttt{Tensor} are
constructed from the same lower-level abstraction (\texttt{Rep[Array[T]]}),
the conversion can be performed at that level.
As an example, Figure~\ref{fig:classifier_generated} (Line 11) illustrates
how records are seamlessly transformed into a tensor (also represented as an array)
by simply assigning the variable.

While this approach might appear to be straightforward in its implementation,
its simplicity belies its effectiveness.
It efficiently eliminates data movement and conversion overheads at system
boundaries—a significant bottleneck in existing tools and a challenging problem as 
highlighted in Section~\ref{section:intro}.
Crucially, it maintains the individual optimizations of each respective system,
ensuring the preservation of individual best-of-breed performance.
In addition, by constructing a single IR for the entire program at the LMS level,
all the IR-level optimizations, including Dead Code Elimination (DCE),
code motion, and loop fusion, are performed globally.
Moreover, this also enables further \textbf{global} optimizations
by downstream general-purpose compilers (GCC or LLVM) during compilation
of the unified generated code.

Our approach does more than simply eradicate the overheads at system
boundaries—it also creates opportunities for further cross-system optimizations
that can considerably enhance performance.
For instance, we can avoid the need to materialize large amounts of intermediate data 
generated by data processing systems for ML models, by fusing the data processing
loops with ML loops.
In the following sections, we delve into several of these key cross optimizations,
all of which significantly improved Flern's end-to-end performance.
A common trait among these implementations is the utilization of high-level
abstractions such as \texttt{Tensor} and \texttt{Buffer} to carry out
cross-system operations and optimizations (Figure~\ref{figure:intro}).

\vspace{-2mm}
\subsection{Optimizing GPU Data Movement}\label{section:gpu_optimizations}

Given the superior efficiency of GPUs in executing tensor computations,
larger ML workloads are typically processed on these units rather than
CPUs.
However, this necessitates the transfer of pertinent data, including model
inputs and weights, from the host (CPU) to the GPU.
In this section, we highlight two key optimizations designed to
enhance the efficiency of this data transfer process, 
and we elucidate how we integrated these improvements into Flern.

By default, any memory allocations made by the host are pageable.
Nonetheless, GPUs cannot directly access this pageable memory.
Thus, when data from such a pageable host array needs to be transferred,
the GPU driver initially allocates a temporary page-locked or ``pinned''
host array.
Subsequently, the data is copied to this freshly allocated pinned array
before being transferred to the GPU memory.
In Flern, we are aware of which data buffers require transfer to the GPU.
This enables us to eliminate the extra data copying step by directly
allocating the corresponding data buffers (used in data manipulation)
in pinned memory.
We add this functionality to Flern by adding a new operator that can
create a \texttt{Buffer} with underlying data (i.e., the C array in
the generated code) directly allocated in pinned memory.
Then, for any \texttt{Buffer} that we invoke \texttt{toTensor()} followed
by a \texttt{toGPU()}, we allocate the original buffer in pinned memory.

Allocating host arrays in pinned memory enables the asynchronous transfer
of data to the GPU.
Typically, when two separate or naively integrated systems are employed for
data manipulation and ML, the GPU remains idle until the data processing phase
concludes.
A primary benefit of generating a single code for the entire task is the
capability to fuse host-to-device data copying operations with data manipulation
operators.
Consequently, data can be transferred to the GPU concurrently with the
production of output from data manipulation.
This allows us to leverage asynchronous memory copying mechanisms 
(e.g., \texttt{cudaMemCpyAsync}), to overlap data copying and data processing.

\vspace{-2mm}
\subsection{Running ML UDFs Efficiently}\label{subsec:efficient_udfs}
	
\begin{figure}[t!]
	\begin{lstlisting}[language=C, basicstyle=\fontsize{7.2}{7}\selectfont\ttfamily\footnotesize, numbers=left, xleftmargin=6mm]
	/* select p, classifier(xs) from r; */
	int main() {
		struct r_record* data = /* load data */
		long record_count = /* data count */
		
		float *w1 = /* classifier weigths*/
		float *b1 = /* classifier bias*/
		float *w2 = /* classifier weigths*/
		
		for(long i = 0; i < record_count; i++) {
			float *tensor = data[i]->xs; // conversion
			// ML Computation
			float *y1 = gemm_kernel(tensor, w1, ...);
			for(int j = 0; j < 4; j++) {
				y1[j] += b1[j];
			}
			float *y2 = gemm_kernel(y1, w2, ...);
			if (*y2 > 0.5) {
				printf("%
			} else {
				printf("%
			}
		}
	}
	\end{lstlisting}
	\caption{Generated code for a simple query that uses an ML classifier
	(variables renamed, some parts elided to improve clarity). Here, \texttt{xs}
	is an array of data and the classifier computes the output by doing two matrix
	multiplications. This code looks exactly like a manually written version of the
	query and performs the combined DB + ML workload.}
	\label{fig:classifier_generated}
\end{figure}

Typically, data processing systems do not natively support ML operators.
Consequently, executing queries involving ML computations often requires
integrating calls to external systems like PyTorch or TensorFlow for ML
functionality.
However, these external functions remain opaque to the data management system,
hampering optimization efforts.
Furthermore, they introduce substantial overheads in the form of data
serialization and invocation, which are required to perform computations
outside the DBMS runtime.
This process often entails significant complexities and inefficiencies
when traversing system boundaries.

With Flern, we can access a comprehensive set of ML operations directly within
the DBMS.
This allows Flern to execute queries involving ML operations as internal
functions.
The advantages of conducting ML computations internally rather than
relying on external systems are manifold.
For instance, these internal ML UDFs are susceptible to optimizations,
facilitating the seamless integration of UDF computations with the rest of the
tasks and enabling efficient optimization during code generation
(e.g., inlining UDF invocations, code motion, etc.).
Furthermore, it allows UDFs to undergo additional global optimizations
at the hands of downstream compilers, like GCC.
Importantly, since all execution takes place within the same runtime,
the overhead associated with invoking the UDF becomes negligible,
leading to significantly improved performance.

Consider the following simple query with a UDF.

\begin{lstlisting}[language=SQL, basicstyle=\fontsize{8}{7}\selectfont\ttfamily]
SELECT x, classifierUDF(y) FROM r;
\end{lstlisting}

Below is a depiction of how an end-user can define such a UDF in Flern using
Scala.
Users can create their UDFs as standard Scala functions, utilizing
functionalities from both Flare (for DB ops) and Lantern (for ML ops).
In the provided code snippet, the user defines a UDF that applies a pre-trained
model to an input value.
The function's signature should align with the number of arguments passed
to the UDF within the SQL query.
When the user registers the UDF, it is added to a collection known as
\texttt{udfMap}, making it readily available for use within the Flern environment.

\begin{lstlisting}[language=Scala, basicstyle=\fontsize{8}{7}\selectfont\ttfamily]

	def classifierUDF(a: Value) = {
	val inputTensor = a.toTensor()
	// run the pre-trained model
	// (can do any arbitrary tensor computation here)
	val output = model(inputTensor)
	output.toValue
}
\end{lstlisting}

In order to incorporate this functionality into Flern, we establish a new
operator, \texttt{ScalaUDF}, which generates code for invoking UDFs as
standard function calls.
Subsequently, the relevant methods in Flare are overridden (as shown below)
to use this newly
introduced UDF operator when UDFs are invoked.
This method retrieves the corresponding function for the \texttt{udfMap} and
executes it with the appropriate parameters.
The generation of code for the function invocation is
handled by the LMS under the hood.
Figure~\ref{fig:classifier_generated} provides a depiction of what the
generated code would look like for the above query.

\begin{lstlisting}[language=Scala, basicstyle=\fontsize{7}{7}\selectfont\ttfamily]
	override def compilerExpr(/* elided */) = {
		case ScalaUDF(function, dataType, children, ...) =>
			// extract the function from udfMap
			val extractedFunc = udfMap(function)
			
			// extract the UDF arguments
			// (need to recursively call compilerExpr because the arguments can be subqueries)
			val values = children map { compileExpr(_)(rec:_*) }
			// call the function 
			extractedFunc(values:_*)
		case _ =>
			super.compilerExpr(/* elided */)
	}
\end{lstlisting}

This approach, however, encounters a significant performance impediment due to
Flare's query execution model, which operates on a single record at a time,
utilizing a modified version of the data-centric
model~\cite{DBLP:conf/sigmod/TahboubER18}.
This is not optimal for executing relatively large UDFs since the overheads
of kernel launches can overwhelm the performance.
As a remedy, we introduced the \texttt{VectorizedUDF} operator that launches
kernels for batches of data, rather than individual instances, thereby amortizing
the kernel launch overhead.
This implementation mirrors the above-described strategy, but instead of
processing single \texttt{Record}s of data, it handles data in batches.

\begin{figure}
	\centering
	\includegraphics[width=0.30\textwidth]{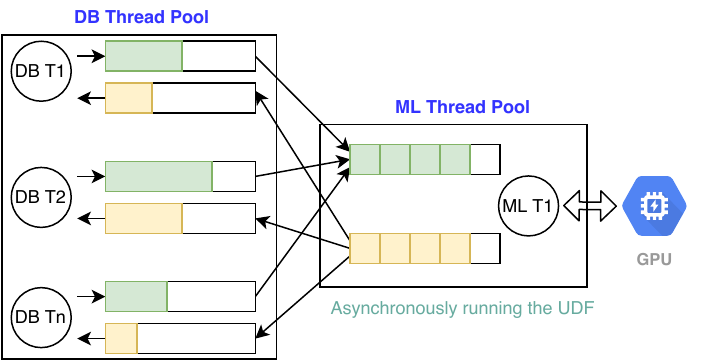}
	\caption{Adding a separate thread pool to run ML UDFs on GPU. These threads
	retrieve records as they are produced, accumulate them and run on the GPU while
	data processing is ongoing.}
	\label{figure:gpu_async_udf}
	\vspace{-4mm}
\end{figure}

While CPUs suffice for running smaller ML models as UDFs, many modern deep
learning models (e.g., a transformer-based sentiment classifier for text data),
demand more computational power, rendering CPU-based computation insufficient.
Consequently, GPU support becomes crucial for executing these UDFs.
We could use the Lantern GPU backend to run \texttt{VectorizedUDF} kernels,
but this approach impedes overall performance as CPU data manipulation is
halted until the output for the current batch is retrieved.
Also, each data processing thread independently and concurrently initiates
GPU kernel launches and data transfers with other CPU threads,
leading to inefficient data transfer and execution.

To address these issues, we have incorporated a new thread pool,
having one thread per GPU, distinct from the data processing threads
(as depicted in Figure~\ref{figure:gpu_async_udf}).
This innovative setup accumulates records as they are generated and then
batches all smaller transfers and kernel invocations together,
thereby efficiently amortizing the overhead.

\vspace{-2mm}
\subsection{Supporting other Popular Front Ends}\label{section:supporting_frontends}

For the effective acceleration of existing end-to-end data science pipelines with
Flern, it's crucial to provide support for widely used front ends,
such as Pandas, Spark, PyTorch, TensorFlow, among others, as highlighted
by~\cite{kagglesurvey}.
Existing tools, like Snek-LMS~\cite{snek}, allow for language virtualization
and source code transformations for Python by implementing a multi-stage
programming framework akin to LMS.
This framework transforms Python code into an IR which can then generate code
in another language.
For example, lightly annotated PyTorch code can be translated into an S-Expression representation
that is then converted into Lantern.

Building on this concept, we could create a translator that generates Flare code
from Python source code.
A more efficient alternative, however, is to utilize the existing front ends of
Spark.
With support for various front-end APIs such as PySpark and Koalas
(which offers a Pandas-like interface), Spark leverages its Catalyst optimizer
to construct optimized query plans
and its core backend for query execution.
Since Flare can operate solely on these optimized query plans built by Spark,
we can extract the query plans generated by these high-level front ends and
employ Flare's runtime for execution.
With Lantern and Flare's ability to support these front ends, 
Flern has the capability to accelerate end-to-end
data science pipelines written using these 
user-friendly APIs, requiring only minimal modifications to the original code.

\vspace{-3mm}
\section{Experimental Evaluation}\label{section:experiments}

We evaluate the performance of Flern in two phases. 
First, we evaluate the impact of
the main global optimizations we have detailed in Section~\ref{section:overview}, such
as minimizing data movement overheads, streamlining UDF execution, and overlapping ML
computations (e.g., GPU) with data manipulations. 
Second, we benchmark Flern's performance against established systems.
A distinct feature of Flern is its ability to rival the performance of top-tier systems in individual workloads. 
As such, we begin by evaluating Flern's standalone performance across a variety of
representative data manipulation and machine learning benchmarks,
comparing it with the leading systems designed for those specific tasks.
It should be noted that, for respective individual workloads, Flern generates code 
that is almost identical to that of Flare and Lantern.
This similarity allows it to replicate previously published experimental
results.
Consequently, we have omitted any results where Flern simply reproduces the
benchmark performance previously published for Flare and Lantern.

For data manipulation workloads, Flern demonstrates the capability to execute the complete
TPC-H benchmark~\cite{tpch}, achieving competitive performance with cutting-edge query
compilers like HyPer~\cite{DBLP:journals/pvldb/Neumann11}, reproducing the results 
in~\cite{DBLP:conf/osdi/EssertelTDBOR18, DBLP:conf/sigmod/TahboubER18}. In this
paper, we include an additional data science-oriented benchmark and compare against 
more baselines. For ML workloads, Flern exhibits the ability to run sophisticated ML 
models such as ResNet~\cite{resnet}, DeepSpeech2~\cite{deepspeech2},
and SqueezeNet~\cite{squeezenet},
matching the performance of leading-edge frameworks like PyTorch and TensorFlow,
reproducing the results in~\cite{DBLP:journals/pacmpl/WangZDWER19, DBLP:conf/nips/WangDWER18}.
In this paper, we include experimental results for a
transformer-based~\cite{DBLP:conf/nips/VaswaniSPUJGKP17} model. 

Subsequently, we analyze the performance of the combined tasks.
This includes relational
queries that combine ML computations and end-to-end data science pipelines where data
manipulation is followed by ML training. 
Our analysis highlights how superficial integrations between task-specific
libraries, such as Spark for data manipulation and PyTorch for deep learning,
can result in added overheads and inhibit potential global optimizations,
leading to suboptimal end-to-end performance.
The primary objective of our experiment section is to substantiate Flern's
comprehensive performance across both types of workloads and to validate its
capacity to match, if not surpass, the best-of-breed performance in
representative benchmarks.

\vspace{-2mm}
\subsection{Benchmark Environment}

For CPU only workloads, we used a NUMA machine with 4 sockets, 24 Intel(R) Xeon(R) Platinum 8168 cores per socket,
and 750GB RAM per socket (3 TB total). For GPU enabled workloads, we conducted experiments on a NUMA machine with
2 sockets, 12 Intel(R) Xeon(R) Gold 6126 cores per socket, and 100GB RAM per socket (200 GB total), a GPU cluster
with four NVIDIA GeForce GTX 1080 Ti 11 GB GPUs. Both servers ran Ubuntu 18.04.4 LTS as the OS.

We use the following versions of libraries/frameworks in the experiments (and the respective versions).
Python 3.7.6, Pandas 1.0.3, PyTorch 1.5.0, TensorFlow 2.2.0, NumPy 1.18.1, Weld 0.4.0, Dask 2.15.0,
CuDF 0.14, Spark 2.4.5, Postgres 12.8, Python-DuckDB 0.8.0, Python-Polars 0.17.12 and GCC 7.5.0.
Each reported result is an average of five runs
after removing the lowest and highest value. Unless otherwise specified, all the tools are running in
multi-threaded mode with the default number of threads which in most cases equal to the number of
logical cores in the system.

\subsection{Effects of Key Optimizations}

\label{section:individiual_optimizations}
\begin{figure}[h]
	\begin{subfigure}{0.25\textwidth}
		\centering
		\includegraphics[width=0.83\textwidth]{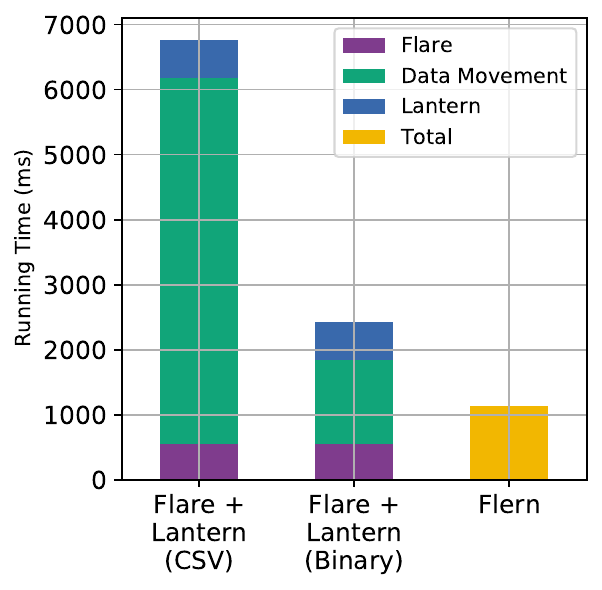}
		\caption{}
		\label{figure:naive_a}
		\vspace{2mm}
	\end{subfigure}%
	~
	\begin{subfigure}{0.25\textwidth}
		\centering
		\includegraphics[width=0.83\textwidth]{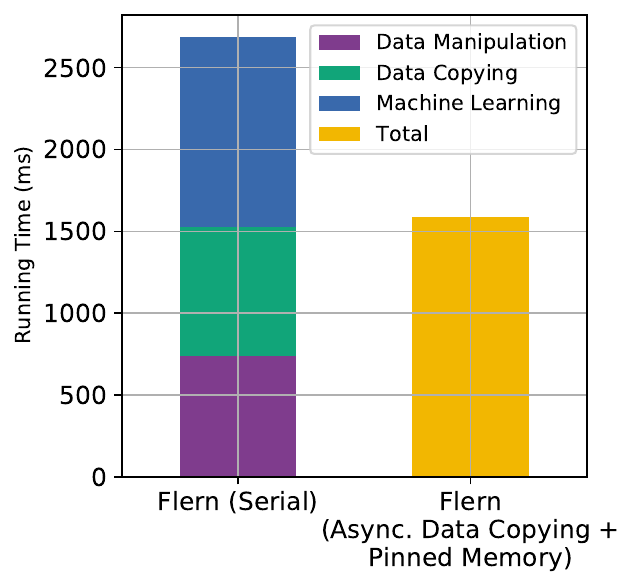}
		\caption{}
		\label{figure:data_movement_2}
		\vspace{2mm}
	\end{subfigure}
	
	\begin{subfigure}{0.25\textwidth}
		\centering
		\includegraphics[width=0.83\textwidth]{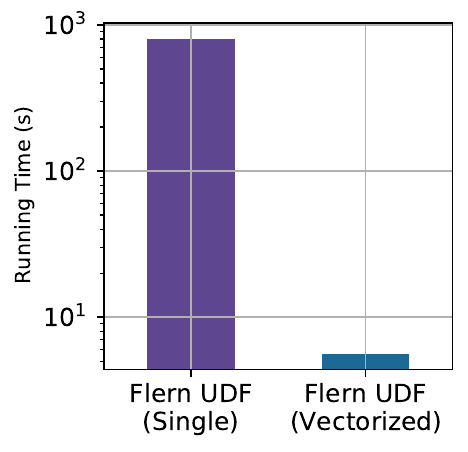}
		\caption{}
		\label{figure:flern_udf_cpu}
	\end{subfigure}%
	~
	\begin{subfigure}{0.25\textwidth}
		\centering
		\includegraphics[width=0.83\textwidth]{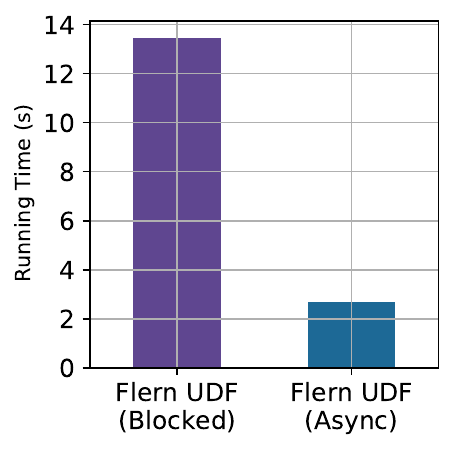}
		\caption{}
		\label{figure:flern_udf_gpu}
	\end{subfigure}
	\caption{Performance impact of global optimizations; (a) using shared data buffers to minimize
	data duplication (CPU), (b) using pinned memory and performing asynchronous data movement to
	GPU (c) vectorizing the UDFs (CPU) (d) running the UDFs asynchronous (in GPU)}
	\label{figure:flern_udf}
	\vspace{-2mm}
\end{figure}

In this section, we evaluate the performance impact of the key optimization steps we have
incorporated.
We used the NYC-Taxi dataset from Kaggle~\cite{nyc_dataset}, which is used for training ML
models to predict the duration of taxi trips based on certain information such as pickup
and drop-off locations, time of day, and so on.
The ML model utilized for this task is a three-layer fully connected neural network,
featuring ReLu activations~\cite{DBLP:journals/corr/abs-1803-08375} between each layer.

\textbf{Minimizing Data Movement Overhead:}
We begin by evaluating the impact of minimizing data movement overhead at system boundaries.
Figure~\ref{figure:naive_a} shows the performance across three scenarios.
One common approach to combining systems is to export the output from one system and
import it from the other.
Such a naive integration between Flare and Lantern, employing a standard data
format (in this case, CSV), results in substantial data movement overhead,
accounting for over 80\% of the execution time.
This is because such formats are not specifically designed for data transfer,
causing both the writer and reader to spend considerable time on
redundant tasks such as formatting and parsing data.
An improvement to this approach is to utilize a binary format, recognized by both systems,
significantly accelerating the data movement process (approximately 2.8 times faster).
On the other hand, as discussed in Section~\ref{subsec:adapting_boundary},
in Flern, high-level abstractions
like Tensors and FlatBuffers translate into native data
structures (i.e., C Arrays) in the generated code, enabling the systems to coexist within
the same memory address space and leverage shared data buffers,
thereby minimizing unnecessary data duplication.
This strategy effectively eradicates the data movement overhead between the systems,
resulting in over a 6x speedup compared to the basic integration.

\textbf{Optimized Host to Device Data Movement:}
In the next experiment, we evaluate the impact of improving the data
movement from CPU memory to GPU memory.
Figure~\ref{figure:data_movement_2} depicts the results for this micro-benchmark.
The first bar represents the sequential case, where the data manipulation
segment of the workload is completed first, followed by the transfer of
the produced data to the GPU.
In this scenario, data transfer consumes nearly 30\% of the total running
time.
With the optimized approach, data is moved in batches as it's being
produced, rather than waiting for the completion of the entire data
manipulation phase.
As a result, data movement is overlapped with data manipulation
computations, effectively increasing efficiency.
Furthermore, extra data copying is eliminated by allocating relevant
buffers in page-locked (i.e., pinned) memory, as explained in
Section~\ref{section:gpu_optimizations}.
By overlapping memory transfers and streamlining the copying process,
we achieve a 40\% reduction in the end-to-end running time.

\textbf{Optimized UDF Execution}
Our next set of experiments evaluates the performance of executing ML models
(or any tensor computations) as internal UDFs in Flern, as detailed in
Section~\ref{subsec:efficient_udfs}.
Figure~\ref{figure:flern_udf_cpu} illustrates the notable performance
improvements of executing UDFs in a vectorized approach,
which yields an execution time more than 140x faster than the serial
approach.
This significant difference can be attributed to the kernel
launch overheads—specifically, the cost of launching the cblas kernels in this case being
amortized across the records in a batch during vectorized execution.
Conversely, in the serial case, each record incurs these costs individually.

Figure~\ref{figure:flern_udf_gpu} shows the execution time for conducting the
same experiment on GPUs.
We have excluded the results for the single-instance version due to its
markedly inferior performance relative to the vectorized versions.
The first bar in Figure~\ref{figure:flern_udf_gpu} represents the scenario
in which the UDF call is blocking and separately invoked by each data
processing thread.
This means that every thread independently initiates data transfers
(between CPU and GPU) and all kernel invocations.
As a result, it leads to a larger number of smaller data transfers and
kernel invocations (due to limited GPU memory), which, adds up to
substantial launch overheads.
Flern addresses this by incorporating a distinct worker thread pool
(as illustrated in Figure~\ref{figure:gpu_async_udf}),
where all data movement and kernel invocations are collectively managed.
In this setting, the dedicated ML threads are capable of moving data and
invoking kernels for large data batches, thereby reducing overall
kernel launch overheads.
This optimization results in a 5x speedup compared to the initial scenario.

\vspace{-2mm}
\subsection{Evaluating Individual Workloads}

In this section, we evaluate Flern's performance across separate individual workloads.
As previously detailed in Section~\ref{section:overview},
Flern preserves all optimizations inherent to the individual systems.
The purpose of the experiments conducted in this section is to verify the
sustained state-of-the-art performance of individual workloads in the absence of 
interactions with other systems.

\begin{figure}
	\centering
	\begin{subfigure}[b]{0.15\textwidth}
		\centering
		\includegraphics[width=\textwidth]{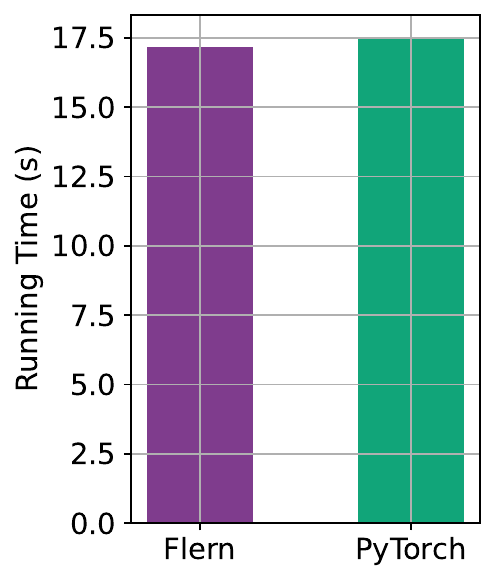}
		\caption{}
		\label{figure:transformer_plot}
	\end{subfigure}%
	~
	\begin{subfigure}[b]{0.33\textwidth}
		\centering
		\includegraphics[width=0.9\textwidth]{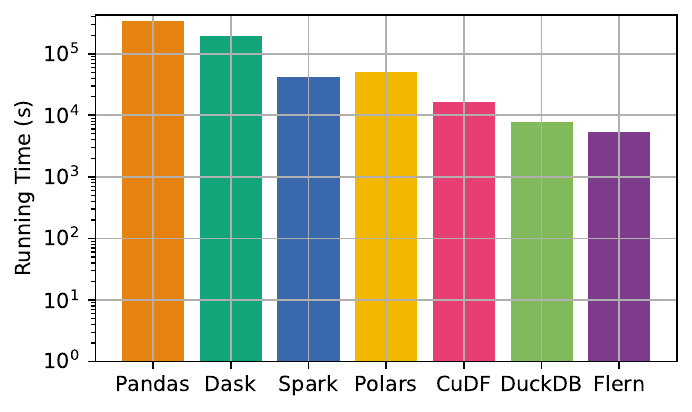}
		\caption{}
		\label{figure:join_query}
	\end{subfigure}
	\caption{Performance comparison for individual workloads (a) ML (b) DB. (a) Per epoch
	 execution time for Transformer model \cite{DBLP:conf/nips/VaswaniSPUJGKP17} on WMT'14 
	 Machine Translation dataset \cite{DBLP:conf/acl/ElliottFSS16} (b) Running Time (s) for 
	 executing multiple (nested) join queries, followed by a \emph{group by} and an 
	 \emph{aggregation} (vertical axis is in log-scale).
	 }
	 \vspace{-4mm}
	
\end{figure}

\subsubsection{Data Manipulation}\label{section:data_manipulation_1}

We evaluate the performance of data manipulation using the Favorita Dataset from Kaggle~\cite{favorita_dataset}.
The dataset consists of six tables representing information about sales of a retail store, 
and the largest table contains 125 million tuples.
We perform a natural join across the six tables, followed by a \emph{group by} and an 
\emph{aggregation} query. This benchmark covers a representative data manipulation 
workload since we perform joins across large relations and performing aggregated queries 
on grouped relations.

Figure~\ref{figure:join_query} shows the results from this benchmark.
Pandas is the go-to framework among data scientists due to its
user-friendly API and popularity~\cite{kagglesurvey}.
However, when used in medium-scale workloads (>1GB), the performance of Pandas becomes inferior due to 
the lack of multi-threaded execution, which results in the longest
execution time for this benchmark.
Dask, in comparison, enhances Pandas' performance by manipulating chunks of Pandas DataFrames
and executing them concurrently~\cite{dask}.
Despite this improvement over Pandas, Dask's execution time remains substantially higher
than the other systems.

Spark, widely acknowledged as the de facto standard for big data processing~\cite{spark_defacto_standard},
is known to exhibit subpar performance when utilized for medium-sized workloads~\cite{DBLP:conf/osdi/EssertelTDBOR18},
yielding lower performance compared to other faster systems.
Despite implementing query compilation techniques that superficially resemble those of Flern,
Spark's performance lags notably in this benchmark.
This gap, as delineated by \citet{DBLP:conf/osdi/EssertelTDBOR18}, can be attributed to the
inherent limitations of Spark's query compilation approach, such as the granularity of
code generation.

CuDF performs the entirety of its data manipulation computations on GPUs with minimal
reliance on CPU.
While CuDF outperforms Spark, it exhibits a longer runtime than Flern.
This gap in performance can be attributed to the overheads related to data spilling
(from GPU to CPU), a consequence of limited GPU memory,
as well as the varying internal algorithms utilized by different platforms
(for example, Hash Join versus Sort-merge Join).
However, for a thorough and accurate comparison, it would be necessary to
conduct an extensive performance analysis, factoring in hardware costs,
given CuDF's primary utilization of GPUs.
Such an analysis, while valuable, lies outside the purview of this paper.

DuckDB~\cite{duckdb} and Polars~\cite{polars}, both relatively recent data analytics frameworks,
are specifically designed for fast single-node, parallel execution.
Much like SQLite~\cite{sqlite2020hipp}, DuckDB is fashioned as an embedded analytical DBMS, 
employing a highly efficient vectorized engine for query execution.
For this benchmark, Flern displays competitive performance compared to these systems,
achieving significant speedups - 9.6x against Polars and 1.44x against DuckDB.
The observed speedup can be attributed to Flern's ability to generate specialized code,
effectively eliminating any runtime interpretive overhead.

Considering Flern's ability to replicate the previously published competitive
performance against leading-edge systems like HyPer~\cite{DBLP:journals/pvldb/Neumann11},
along with its impressive results on
new benchmarks, it can be confidently stated that Flern is capable of achieving performance
on par with the state-of-the-art systems for data manipulation workloads.

\subsubsection{Machine Learning}
In the preceding section, we demonstrated how Flern significantly outperforms 
the performance of many commonly utilized data processing libraries and systems
within the data science community.
In this section, we shift our focus to evaluate Flern's capabilities when executing
state-of-the-art deep learning models.
Specifically, we trained a transformer-based machine translation model~\cite{DBLP:conf/nips/VaswaniSPUJGKP17}
on the WMT'14 Multimodal Translation dataset~\cite{DBLP:conf/acl/ElliottFSS16}.
The model mirrors the configuration of the original Transformer-base 
model~\cite{DBLP:conf/nips/VaswaniSPUJGKP17},
employing 6 encoder and decoder layers and an embedding size of 256,
while parameters are updated using the Adagrad optimizer.

Figure~\ref{figure:transformer_plot} provides a comparison of the per-iteration
running time for Flern against PyTorch.
Given the similarity of the internal CUDA kernels used by both systems,
their performance is almost equivalent.
This experiment's key finding illustrates that Flern can execute these
cutting-edge deep learning models with competitive performance,
a capability that has not been demonstrated by prior works developing common
intermediate layer integrations~\cite{DBLP:conf/cidr/PalkarTSSAZ17, DBLP:conf/ecoop/SujeethRBLCPWPJOO13}.

\vspace{-2mm}
\subsection{Evaluating Combined Workloads}

To assess the performance of Flern in combined workloads, we present three sets of 
experiments. 
Firstly, we execute an augmented User-Defined Function (UDF) that incorporates a tensor
computation within it.
Secondly, a set of experiments where a complete ML model is used as a UDF, 
executed on both CPU and GPU.
Lastly, we implement an end-to-end pipeline where an ML model is trained 
using the data obtained via a relational query.

\subsubsection{Enhanced UDFs}

In this section, we evaluate the performance of Flern in comparison with Weld,
a system that bears superficial similarities to our own.
We selected a benchmark from one of Weld's openly available implementations in
which a scalar value (a crime index) is computed via dot product involving a
chosen subset of columns (crime-related data) and a predefined constant vector.
This benchmark showcases a use case where multiple libraries are jointly employed.
Specifically, in the case of Weld, the data manipulation portion is handled by Grizzly,
a Pandas accelerator based on Weld, and the dot product is handled by Weld-NumPy. 
Similarly, Flern utilizes Lantern for dot product operations.

\begin{figure}
	\centering
	\begin{subfigure}{0.27\textwidth}
		\centering
		\includegraphics[width=0.7\textwidth]{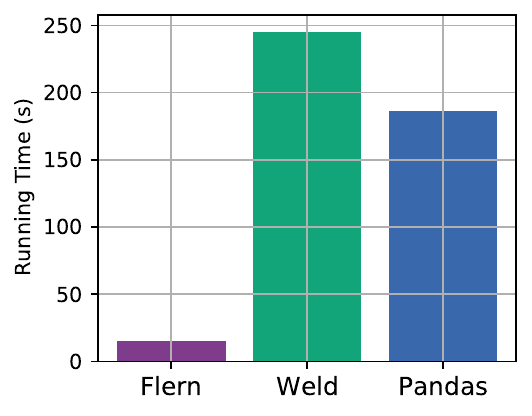}
		\caption{}
		\label{figure:weldbench1}
	\end{subfigure}%
	~
	\begin{subfigure}{0.22\textwidth}
		\centering
		\includegraphics[width=0.7\textwidth]{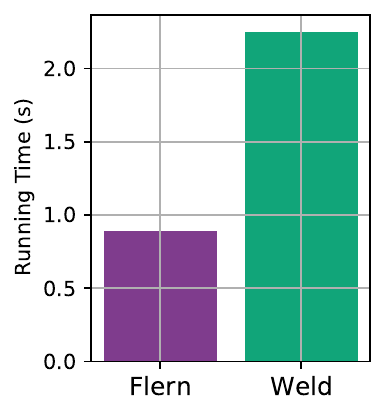}
		\caption{}
		\label{figure:weldbench2}
	\end{subfigure}
	\caption{Execution time for performing a dot product as a UDF in an SQL query. 
	(a) shows the end-to-end execution time whereas (b) compares the execution
	time of the generated code of Weld and Flern.}
	\vspace{-4mm}
\end{figure}

Figure~\ref{figure:weldbench1} shows the end-to-end execution time, including 
time for data loading, code generation, and compilation (for both Weld and Flern).
Intriguingly, Weld's end-to-end time surpasses that of Pandas.
This can be primarily attributed to Weld's reliance on Pandas for data loading
and the time expended on marshaling and demarshaling data between the host
language and Weld's runtime.
Figure~\ref{figure:weldbench2} provides a glimpse into the actual runtime of the
code generated by Weld and Flern, excluding data loading and any other
communication overheads.
Weld's generated code has been optimized through several passes,
including loop fusion, vectorization, loop unrolling, and others,
as well as LLVM level optimizations~\cite{DBLP:journals/pvldb/PalkarTNTPNSSPA18}.
Similarly, Flern's generated code undergoes a multitude of optimizations at various
levels, from domain-specific ones to code specialization and
low-level compiler optimizations.
Notably, Flern's resultant code exhibits a 2.5x speedup compared to Weld.
This suggests that while both systems construct a common IR across systems and
generate low-level code, the methodology employed to architect such IRs
can have impacts on the final performance.

\subsubsection{Running ML Classifiers as UDFs}\label{section:eval_ml_udf}

In this section, we assess the performance of executing ML classifiers as
UDFs in relational queries, comparing Flern's performance with Spark, Postgres,
and DuckDB with which PyTorch serves as the ML framework.
We note that the publicly available Weld implementation does not support the workloads
in this section (and Section~\ref{section:e2e_training}).

In the case of Spark, it employs Apache Arrow~\cite{arrow}, a common in-memory columnar data
format for data transition between JVM and Python environments.
Arrow minimizes data serialization and transfer costs, thereby cutting down the
communication overhead across system boundaries.
This strategy considerably outpaces the conventional method of executing the UDF
as an external function on a per-record basis (i.e., scalar UDF),
with serialization/deserialization at system boundaries 
(the numbers for this case are not reported).
For Postgres, we employed the PL/Python plugin and implemented the ML operations
using PyTorch. 
We have adopted a similar approach for DuckDB as Spark, 
leveraging Arrow to efficiently transfer data to PyTorch 
in a vectorized manner.
Specifically, we used the provided utilities to convert the 
relation into batches of records, which were then streamed to 
PyTorch for performing ML inference.

This experiment reuses the same dataset and model from 
Section~\ref{section:individiual_optimizations} for the CPU scenario
and employs a Transformer-based sentiment classifier on the Yelp Reviews
Dataset~\cite{yelp_dataset} (1 million reviews) for the GPU scenario.
Specifically, for sentiment analysis, we calculate the aggregate number
of positive and negative reviews based on restaurant star ratings
(the query is displayed below).
The sentiment classifier contains word and positional 
embeddings~\cite{DBLP:conf/nips/VaswaniSPUJGKP17}, six encoder layers,
succeeded by a three-layer fully connected layer with 
ReLu~\cite{DBLP:journals/corr/abs-1803-08375} activations.

\begin{lstlisting}[language=SQL,basicstyle=\fontsize{6.4}{1}\selectfont\ttfamily]
SELECT 
	stars, 
	Sum(CASE WHEN sentiment < 0.5 THEN 1 ELSE 0 END) AS negative, 
	Sum(CASE WHEN sentiment >= 0.5 THEN 1 ELSE 0 END) AS positive 
FROM
	(SELECT stars, classifier(review) AS sentiment FROM review NATURAL JOIN business) 
GROUP BY stars 
\end{lstlisting}

\begin{figure}[t!]
	\centering
	\begin{subfigure}[b]{0.25\textwidth}
		\centering
		\includegraphics[width=0.95\textwidth]{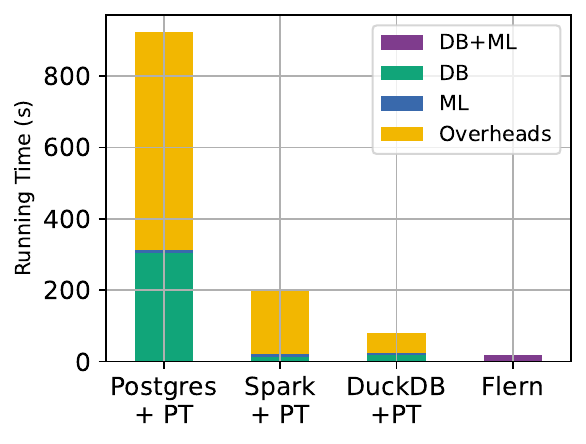}
		\caption{}
		\label{figure:spark_udf_cpu}
	\end{subfigure}%
	~
	\centering
	\begin{subfigure}[b]{0.25\textwidth}
		\centering
		\includegraphics[width=0.95\textwidth]{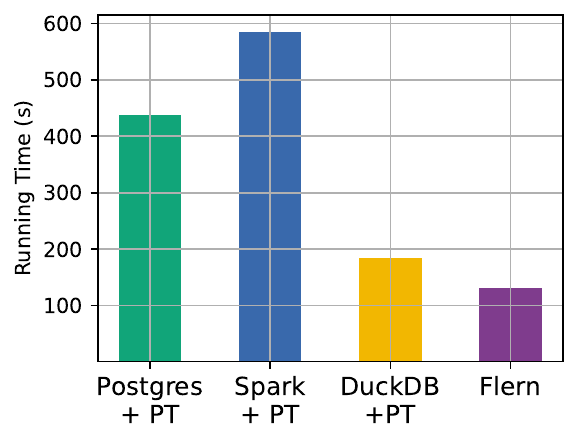}
		\caption{}
		\label{figure:spark_udf_gpu}
	\end{subfigure}
\caption{Performance comparison for (a) Running a 3-layer Neural Network 
(Regression) UDF on CPU (b) Running a Transformer-based Sentiment Classifier UDF on GPU }
\vspace{-3mm}
\end{figure}

Figure~\ref{figure:spark_udf_cpu} shows the execution time for the CPU scenario,
in which Flern showcases speedups exceeding 50x against Postgres, approximately
12x over Spark, and around 4x over DuckDB.
We conducted numerous micro-benchmarks to uncover the reason behind this substantial
speedup.
For the Postgres+PT, Spark+PT, and DuckDB+PT scenarios, we ran the corresponding
workloads on each individual system (i.e., Postgres, Spark, DuckDB, and PyTorch separately)
to gauge the extent of overheads introduced due to the integration.
Figure~\ref{figure:spark_udf_cpu} emphasizes the time portion allocated for
actual computation.

These overheads primarily stem from data copying and format conversion overheads
occurring at system boundaries.
Specifically, for Spark, this comprises the cost of converting Spark DataFrames to
Arrow, transitioning data from one execution environment to another,
converting Arrow data to Pandas DataFrames, and so on.
Similar data copying happens in the Postgres and DuckDB scenarios as well.
Such overheads account for more than 90\% of the total execution time for
Postgres and Spark, and exceed 70\% for DuckDB.
As Flern generates a unified piece of low-level code with the UDFs inlined
and fused with the data processing loops, none of these overheads exist.
Furthermore, these UDFs are transparent to the compiler and undergo further
global optimizations.

Figure~\ref{figure:spark_udf_gpu} presents the execution time for the GPU scenario.
Interestingly, the Postgres implementation runs faster than Spark.
This is likely due to a lack of coordination among the Spark workers when executing
computations on the GPU.
Specifically, each worker independently dispatches computations to the GPU,
leading to multiple workers launching kernels for smaller data batches
(constrained by the limited GPU memory) simultaneously.
This results in an increase in kernel launches, smaller but more frequent data
transfers between host and device, and consequently, an increase in launch overheads,
which becomes the dominant factor.
As our micro-benchmarks in Section~\ref{section:individiual_optimizations} indicated,
such lack of coordination can significantly impede performance.
Conversely, in the case of Postgres, a single PyTorch worker is initiated by the DB
to handle all GPU computations, thereby minimizing such kernel launch overheads.
However, this approach has the drawback of requiring the complete set of records to
be ready before invoking the ML computation. 
DuckDB outperforms Spark due to its highly efficient vectorized query engine
tuned for single node workloads.
As outlined in Section~\ref{section:gpu_optimizations} (Figure~\ref{figure:gpu_async_udf}),
Flern employs a separate thread pool that batches ML computations from data processing workers,
which mitigates the kernel launch overheads.
This, combined with the elimination of system boundary crossing overheads,
makes Flern notably faster than the other systems.

It's also important to highlight that while the cross-system communication overhead
is confined to a single occurrence for the two queries chosen for our experiments
in this section, scenarios may arise where this communication recurs multiple times
within a single query, multiplying the impact of aforementioned cross-system overheads.
Seamless integration in the style of Flern would be even more advantageous in such scenarios.

\subsubsection{End-to-end Training}\label{section:e2e_training}

\begin{figure}
	\centering
	\includegraphics[width=0.35\textwidth]{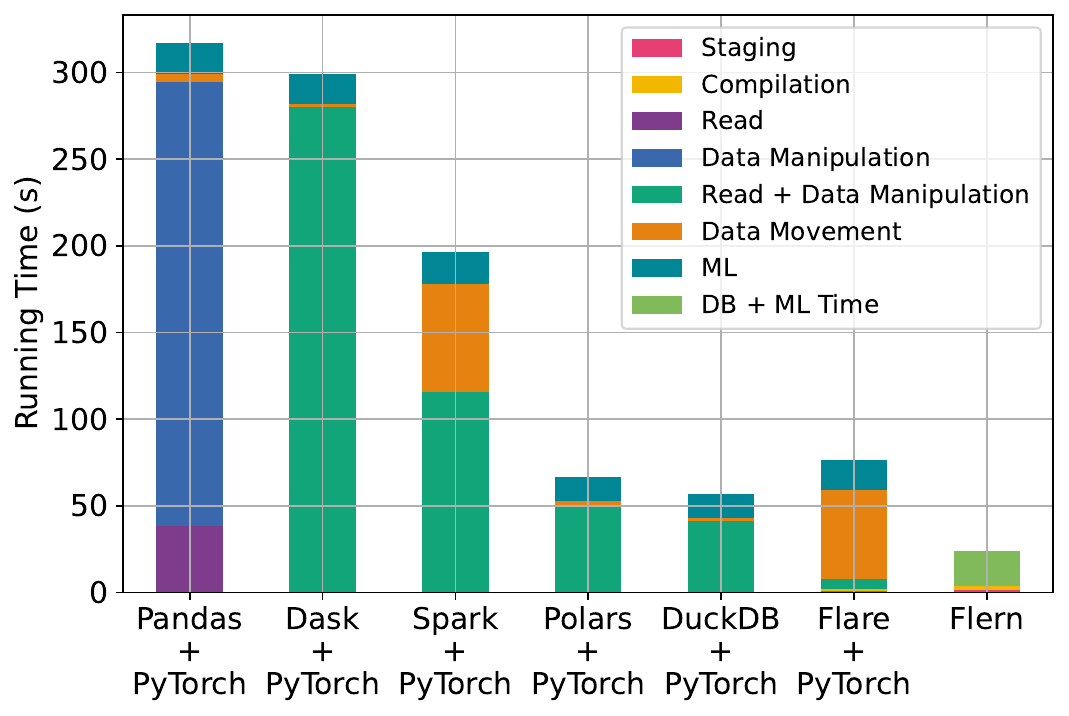}
	\caption{
	Execution time (ms) for an end-to-end ML training pipeline consisting of
	querying a relational source, performing a join, and training an ML model.
	Each bar segment in the chart denotes the time duration
	dedicated to specific subtasks (such as data reading and
	performing joins, etc.) in each case.
	}
	\label{figure:e2e_training}

	\vspace{-4mm}
	
\end{figure}

In this section, we assess the performance of Flern in the context of training an
ML model using data obtained through a relational query.
We use the Favorita dataset~\cite{favorita_dataset} from Kaggle, 
identical to the one used in Section~\ref{section:data_manipulation_1}.
The task involves training a model to predict sales numbers accurately for a
large grocery chain.
This particular task has also been utilized as a benchmark in previous related 
studies~\cite{ifaq, factml_lmfao}.
The dataset contains multiple relational tables, including related information such
as daily oil prices, item-specific data (for instance, whether an item was promoted),
transaction histories, and more.
Similar to \citet{ifaq}, we initially conduct a natural join operation across
all the relations, subsequently selecting all numerical columns.
We then train a neural network comprising three layers, with 
ReLU~\cite{DBLP:journals/corr/abs-1803-08375} activations used in the first two layers.
We compute the Mean Squared Error (MSE) loss, calculate gradients concerning the
loss, and leverage Stochastic Gradient Descent (SGD) to optimize the model parameters.
For this benchmark, the ML computations are run on the GPU.
We run this benchmark on a selection of the most widely used and high-performance
systems in the DB and ML domains.

The results of this benchmark are summarized in Figure~\ref{figure:e2e_training}.
The performance mirrors the pattern observed in Section~\ref{section:data_manipulation_1} for 
data manipulation benchmarks, attributable to the same identified limitations.
CuDF was unable to complete the experiment due to insufficient GPU memory. 
Given that Flare performs runtime native code generation, there is no \emph{direct}
way of integrating it with a library like PyTorch, making data movement a bottleneck
in this context.
In contrast, Flern, which establishes an end-to-end compiled data path for executing the combined workload,
outperforms all other approaches.
Flern achieves a speedup exceeding 13x compared to the original Pandas version and
manages around a 3.2x speedup over a basic integration between Flare and PyTorch.
DuckDB and Polars, with their highly efficient parallel query engines,
come closest to achieving the performance of Flern. However, Flern still outperforms
them by a factor of approximately 3x, primarily due to its tight integration with the ML
system.

\vspace{-3mm}
\section{Related Work}\label{section:related}

Weld \cite{DBLP:conf/cidr/PalkarTSSAZ17, DBLP:journals/pvldb/PalkarTNTPNSSPA18} is a common
runtime specifically targeting data-intensive applications with a main focus on physical data
movement optimizations. Weld provides a unified runtime for multiple library front ends 
such as Pandas~\cite{pandas} and NumPy~\cite{DBLP:journals/cse/WaltCV11}
by implementing the corresponding operations using the Weld API that constructs
Weld IR fragments.
Delite~\cite{DBLP:journals/tecs/SujeethBLRCOO14, DBLP:conf/ecoop/SujeethRBLCPWPJOO13} is a 
compiler framework for implementing embedded domain-specific languages (DSLs).
The Distributed Multiloop Language (DMLL)~\cite{DBLP:conf/cgo/BrownLRSSAO16} 
performs cross optimizations on data processing and ML operations. 
Lara~\cite{lara_vldb_19,lara_short} is a DSL that supports both collection processing
and ML operations, and accompanies an associated IR that represents the combined workload.
Despite these systems' capability to enhance performance in combined workloads,
their performance either falls short in isolated tasks, failing to reach state-of-the-art
or they lack the comprehensive functionality required to run extensive benchmark suites
such as full TPC-H or modern ML models. Moreover, these systems necessitate the ground-up construction of individual
systems or substantial modifications to pre-existing systems, leading to
increased developmental complexity and effort.

Split Annotations~\cite{DBLP:conf/sosp/PalkarZ19} identifies these limitations and 
presents an approach that can combine existing systems without modification. 
Specifically, they treat library functions as black boxes and adds a cross-function
data pipelining layer to curb data movement overheads. 
Nonetheless, this approach sacrifices the benefits of building common IRs and performing
code generation, such as cross-system operator fusion.

Numerous prior works have investigated the approaches to integrate ML computations
into DB systems to circumvent the expensive data movement
overheads~\cite{dbml_db4ml, madlib, bismarck, dbml_gpu, raven}.
MADLib~\cite{madlib} and Bismarck~\cite{bismarck}, for instance, 
encapsulate ML computations as user-defined functions (UDFs) within the 
DBMS.
Furthermore, factorized ML approaches efficiently perform ML operations on multi-table
data by pushing ML computations to the normalized relations, thus eliminating the
necessity to materialize extensive join
results~\cite{factml_arun3, factml_arun4, factml_f,factml_f1, factml_lmfao, factml_main, factml_morpheus_1, factml_morpheus_2}.
However, these approaches tend to be algorithm-specific and either do not support the
full range of modern ML models or lean on external systems like
TensorFlow~\cite{DBLP:conf/osdi/AbadiBCCDDDGIIK16} for more complex
models~\cite{madlib_dl}.

Raven~\cite{raven,raven_sigmod} extends ONNX IR~\cite{onnx} by
incorporating relational algebra operations, thus creating a unified
IR for the combined workload and facilitating cross-optimizations.
Similarly, Umbra~\cite{umbra2020_cidr,umbra_dl} integrates compiled
TorchScript~\cite{torchscript, DBLP:conf/nips/PaszkeGMLBCKLGA19} modules as
relational algebra operations in the query plan. 
Despite the considerable flexibility these approaches provide, 
they still rely on disparate systems for actual execution,
which leads to overheads at the interfaces between these systems.

\citet{klabe2023exploration} delve into a variety of approaches for in-DB ML.
Our approach mostly aligns with their proposed `native operator' approach,
which they perceive as the most performance-optimized route.
Although they postulate that this path might necessitate substantial engineering
effort, Flern actualizes this design using an existing ML framework,
thereby circumventing the need for implementing these operators from scratch.

Many DBMS and cloud vendors offer in-DB ML support. 
These include, Google BigQuery ML~\cite{biqueryml},
Microsoft SQL Server ML~\cite{sqlserverml},
Redshift-ML~\cite{redshiftml},
Postgres (via PL/Python~\cite{plpython}),
Oracle~\cite{oracleml}, Vertica-ML~\cite{fard2020vertica}.
Their main focus is to provide a user interface designed for handling queries
that incorporate ML computations, although the execution of these computations
is often delegated to external systems~\cite{klabe2023exploration}.

\vspace{-3mm}
\section{Conclusions}\label{section:conclusions}
Modern data analytics workloads frequently intertwine DB and ML computations.
Yet, the integration of ML capabilities with DBMSs (or vice versa) poses significant
challenges due to the contrasting nature of these workloads.
While individual ML systems and DBMSs are optimized for their specific tasks,
the performance often diminishes significantly when these systems are
integrated naively.
A promising strategy to mitigate these overheads involves architecting
individual systems first and then integrating them at a common intermediate layer,
facilitating the generation of a unified executable for the combined workload.

This paper introduces Flern, the first such intermediate layer that achieves best-of-breed
performance in both individual \emph{and} combined workloads.
Our approach, based on generative programming, demonstrates how common
intermediate layers enable efficient integration between existing systems without
incurring substantial re-engineering costs.
We posit that this paper outlines a principled methodology to address this
challenge in general: architect systems based on generative 
programming so that they can be adapted more effectively and at a lower engineering 
cost.

\bibliographystyle{ACM-Reference-Format}
\bibliography{files/references}

\end{document}